# 100 G Data Center Interconnections with Silicon Dual-Drive Mach-Zehnder Modulator and Direct Detection

Xiaoke Ruan, Lei Zhang, Fan Yang, Yixiao Zhu and Fan Zhang, *Senior Member, IEEE, Senior Member, OSA*

*Abstract*—In this paper, we experimentally demonstrate that a silicon dual-drive Mach–Zehnder modulator (DD-MZM) has great potential for next-generation data center interconnections (DCIs). For intra-data center interconnections, 120 Gb/s Nyquist 4-ary pulse amplitude modulation (PAM-4) signal is successfully generated with a silicon DD-MZM operating at C-band and transmitted over 2 km standard single-mode fiber (SSMF) with a bit error rate (BER) of $5.55 \times 10^{-4}$. For inter-data center interconnections, single sideband (SSB) modulation is chosen to avoid power fading caused by fiber chromatic dispersion and square-law detection. We report the generation and transmission of 112 Gb/s Nyquist SSB PAM-4 signal by using the same silicon DD-MZM and Kramers-Kronig (KK) direct detection. A two-tap digital post filter and maximum likelihood sequence detection (MLSD) are applied to compensate for the limited system bandwidth. After 80 km SSMF transmission, the BER is $2.46 \times 10^{-3}$ that is below the 7% HD-FEC threshold of $3.8 \times 10^{-3}$. To the best of our knowledge, our work reports the highest single-lane bitrate of 80 km SSB transmission based on a silicon DD-MZM. Our study also shows the feasibility of silicon photonic modulator for DCI applications in the future.

## I. INTRODUCTION

Driven by the ever-growing demand for high-capacity services such as social media, video streaming, and cloud computing, high-speed optical transmission solutions to data center interconnection (DCIs) based on direct detection have been extensively studied owing to their relatively low complexity and low cost compared to coherent detection. For intra data center interconnections (intra-DCIs) with transmission distance of 500 m, 2 km and 10 km, the four-level pulse amplitude modulation (PAM-4) has been adopted by IEEE802.3bs Task Force for 400GbE multi-channel transmission [1]. PAM-4 is advantageous for transmission over band-limited channels and allows the use of lower bandwidth electro-optic components. Recently, Huawei Technologies reported single-lane 180 Gb/s PAM-4 signal transmission over 2 km single-mode fiber (SSMF) by using a commercial modulator [2]. However, for inter data center interconnections (inter-DCIs) with optical links up to 40-80 km, PAM-4 is not applicable anymore since fiber chromatic dispersion (CD) will introduce severe power fading distortion to conventional double sideband (DSB) systems [3] due to square-law direct detection. To overcome the CD induced spectrally selective fading in DSB systems, vestigial sideband (VSB) modulation or single sideband (SSB) modulation can be applied to direct detection systems. For VSB modulation, a conventional DSB signal is first generated at the transmitter side and then a sharp-edged optical bandpass filter (OBPF) is applied to suppress half of the spectrum [4]. However, VSB modulation is not suitable for practical application since a sharp-edged optical filter will significantly increase the cost of the transceiver and the limited sharpness of the filter edge usually leads to a non-ideal SSB spectrum which will degrade the system performance. Instead of optically generated VSB, the directly generated SSB signals through digital Hilbert transform by using two digital-to-analog converter (DAC) channels with an IQ modulator [5,6] or a dual-drive intensity modulator [7,8] have been widely demonstrated, which can offer a much better optical sideband suppression than VSB signals. The signal-signal beat interference (SSBI) is a main impairment due to square-law detection of the SSB signal, which can be mitigated with the iterative SSBI cancellation algorithm [5] or Kramers-Kronig (KK) receiver [9].

For DCI applications, the expensive discrete devices employed in commercial communication systems may not be suitable. In contrast, silicon photonic devices have emerged as a candidate solution owing to their small footprint, high integration density, and low cost. In specific, silicon modulators have received ever-increasing attentions from the research and industrial community. Among multiple kinds of silicon modulators, the depletion-type of Mach-Zehnder modulator (MZM) [10] is the most common for practical use and has been applied in various modulation schemes. For intensity modulation, McGill University has demonstrated 112 Gb/s PAM-4 transmission over 2 km SSMF with bit error rate (BER) below hard-decision forward error-correction (HD-FEC) threshold of $3.8 \times 10^{-3}$ by using a silicon single-drive modulator in 2015 [11]. Recently, Huazhong University of Science and Technology reported 128 Gb/s PAM-4 transmission over 2 km by using a silicon single-drive modulator with a BER below FEC threshold of $2 \times 10^{-2}$ [12].

This work was supported by National Natural Science Foundation of China (No. 61535002 and No. 61475004).
Xiaoke Ruan, Lei Zhang, Fan Yang, Yixiao Zhu and Fan Zhang are with the State Key Laboratory of Advanced Optical Communication Systems and Networks, Peking University, Beijing, 100871, China (e-mail: ruanxiaoke@pku.edu.cn; zhang_lei@pku.edu.cn; fianfian@pku.edu.cn; yxzhu@pku.edu.cn; fzhang@pku.edu.cn).

For IQ modulation, in 2016, Huawei Technologies have successfully demonstrated a 80×453.2 Gb/s polarization division multiplexing (PDM) orthogonal frequency-division multiplexing (OFDM) dense wavelength division multiplexing (DWDM) signals transmission over 480 km and the single-lane net bit rate is about 300 Gb/s by using a silicon single-drive IQ modulator [13]. Recently, Université Laval has demonstrated single polarization 350 Gb/s 32-ary quadrature amplitude modulation (32-QAM) with a net rate of 291 Gb/s by using a silicon single-drive IQ modulator [14]. For SSB modulation, Huawei Technologies demonstrated 56 Gb/s SSB discrete multi-tone (DMT) transmission over 320 km by using a silicon single-drive IQ modulator in 2015 [13]. In 2017, Peking University achieved 40 Gb/s SSB 16-QAM transmission over 160 km by using a silicon dual-drive intensity modulator and 50 Gb/s SSB 16-QAM transmission over 320 km by using a silicon dual-drive IQ modulator, respectively, both with electrical packaging [15]. For VSB modulation, McGill University has recently reported 112 Gb/s PAM-4 transmission over 40 km with a BER below the FEC threshold of $3.8 \times 10^{-3}$ and 120 Gb/s PAM-4 transmission over 80 km with a BER below the FEC threshold of $2 \times 10^{-2}$ by using a silicon multi-electrode Mach-Zehnder modulator to generate the original DSB real signal and a sharp-edged OBPF to remove the unwanted sideband [16]. All of the above mentioned silicon modulators are designed for C-band.

In this paper, we experimentally demonstrate that a simply structured silicon dual-drive Mach–Zehnder modulator (DD-MZM) is capable for next-generation DCI applications. For intra-DCIs, we generate 120 Gb/s Nyquist PAM-4 signal based on a silicon DD-MZM operating at C-band. After transmission over 2 km SSMF, the received BER of $5.55 \times 10^{-4}$ is well below the 7% HD-FEC threshold of $3.8 \times 10^{-3}$ and the net bitrate of the system is about 109 Gb/s. For inter-DCIs, we generate 112 Gb/s Nyquist SSB PAM-4 signal to overcome the CD induced spectrally selective fading. After 80 km SSMF transmission, the SSB signal is reconstructed with KK direct detection receiver. With the help of a two-tap digital post filter and maximum likelihood sequence detection (MLSD) [17] to compensate for the limited system bandwidth, the measured BER after transmission is $2.46 \times 10^{-3}$ that is below the 7% HD-FEC threshold of $3.8 \times 10^{-3}$ and the net bitrate achieves about 102 Gb/s. To the best of our knowledge, our study reports the highest single-lane bitrate for C-band 80 km SSMF SSB signal transmission based on a silicon DD-MZM and direct detection.

## II. DEVICE CHARACTERIZATION

Fig. 1(a) shows the structure of the silicon traveling-wave Mach-Zehnder modulator (TWMZM) fabricated through IMEC's silicon photonics ISIPP50G technology. The intended device is highlighted by the white dashed line. As shown in Fig. 1(b), on the left side of the signal lines, a group of GSGSG pads are built to apply the bias voltage and the high-speed driving signal through the electrical RF probe. As shown in Fig. 1(c), the light is coupled in and out of the waveguide device by fiber-to-chip grating couplers with an insertion loss of ~5 dB/coupler. Each arm of the TWMZM contains a phase modulator of 2.5 mm length. The phase modulators operate via the plasma dispersion effect, where the depletion of free carriers from a reverse biased p-n junction embedded in the waveguide causes a phase shift of the propagating light. The $V_\pi L$ of each modulation arm is about 1.73 V-cm. In order to match closely with the characteristic impedance of the transmission line, each signal line is terminated with a 25 Ω on-chip resistor, which, as shown in Fig. 1(d), is established by two parallel 50 Ω n-doped silicon slabs between the ground and the signal. The waveguides of the two arms are intentionally designed with 20 um length difference away from the modulation region, which allows the modulator's operation point to be adjusted through wavelength tuning. Localized heating elements are positioned in each Mach-Zehnder arm, which can also adjust the operation point. For simplicity, we choose tuning wavelength for operation point adjustment in our experiments.

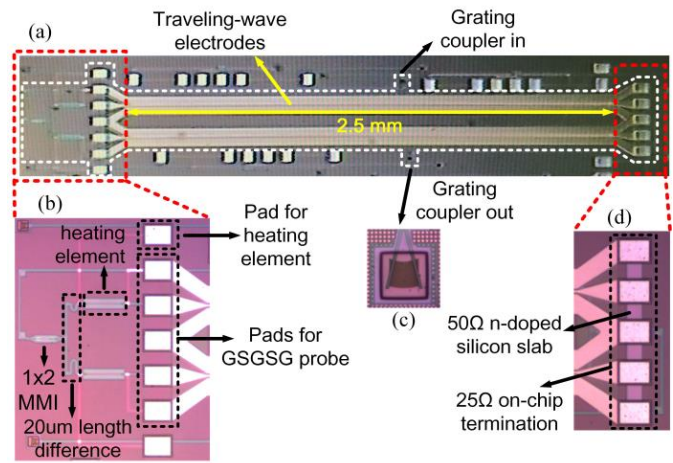

Fig. 1. (a) Micrograph of the TWMZM. (b) Zoom-in view of the left part. (c) Zoom-in view of the grating coupler. (d) Zoom-in view of the right part.

Fig. 2(a) shows the measured electro-optic response of the MZM under different reverse bias voltages. With the reverse bias voltage varying from 0 V to 3 V, the 3-dB bandwidth of the MZM gradually increases from 14.7 GHz to 21.4 GHz and then stays almost unchanged. The increased bandwidth is mainly due to the reduction of lateral p-n junction capacitance, which is analog to the reduction of RC time constant. In the following experiments in Section III and IV, we choose 2 V as the reverse bias. Fig. 2(b) shows the measured optical spectra of the MZM. The results have been normalized by the response of a reference waveguide to eliminate the influence of the grating coupler. By simultaneously changing the reverse bias voltages on both arms from 0 V to 3 V, the spectra have obvious red shift since the two arms have slightly different phase shift performance, which is mainly due to mask alignment errors. In the following experiments, in order to make the modulator works at the quadrature point, we set the laser source operating at 1552.9 nm that is illustrated as the yellow circle in Fig. 2(b).

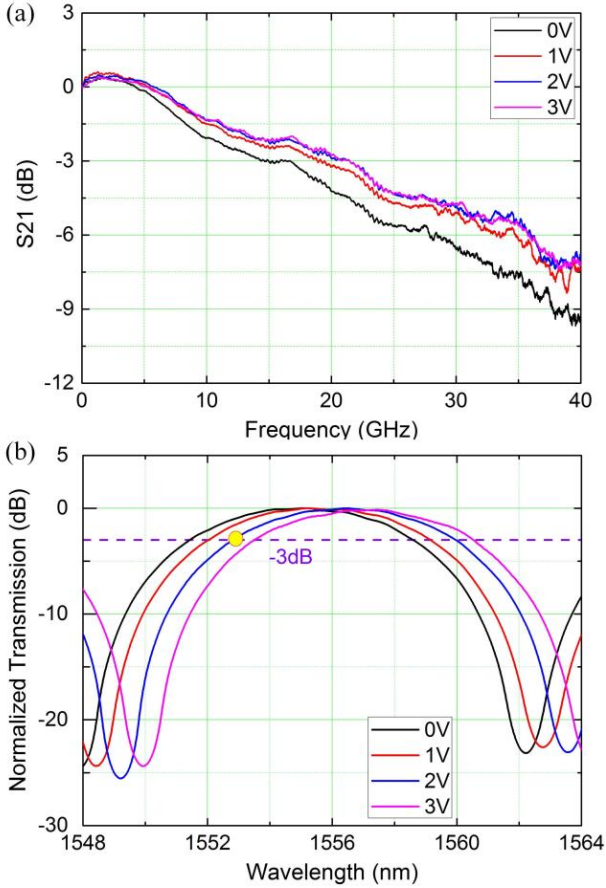

Fig. 2. (a) Electro-optic response of the TWMZM at various reverse bias voltages. (b) Optical spectra of the TWMZM when both arms under the same reverse bias voltages.

## III. PAM-4 Transmission

### A. Setup and DSP

First we generate PAM-4 signals by using the above DD-MZM to meet the demand for short-reach intra-DCIs. The experimental setup is shown in Fig. 3(a). At the transmitter, the wavelength of the external cavity laser (ECL) has been precisely tuned to 1552.9 nm that is illustrated as the yellow circle in Fig. 2(b) to make the silicon DD-MZM works at the quadrature point. A polarization controller (PC) is applied to adjust the polarization state of the light coupling to the chip. The on-chip silicon DD-MZM is fed by an arbitrary waveform generator (Keysight M8195A) operating at 65 GSa/s to generate 60 Gbaud PAM-4 Nyquist-shaped signal. After a polarization-maintaining erbium-doped optical fiber amplifier (PM-EDFA), the optical signal is launched into a 2 km SSMF link. At the receiver, the optical signal is first detected by a photodiode (PD) and then amplified by an electrical amplifier both with 50 GHz bandwidth. Finally, the electrical signal is sampled by a digital storage oscilloscope (DSO) operating at 160 GSa/s.

Fig. 3(b) shows the digital signal processing (DSP) diagrams. At the transmitter side, the bit stream is mapped to PAM-4 format first. After 13 times up-sampling, the signal is digitally shaped using root raise cosine (RRC) filter with a roll-off factor of 0.01. Then the signal is 12 times down-sampled. Hence the baud rate is 60 Gbaud when the arbitrary waveform generator (AWG) is operated at 65 GSa/s. For the receiver DSP, the signal is firstly re-sampled to 4 samples per-symbol. After the matched RRC filter and synchronization, the signal is equalized with a Ts/4 (Ts means the symbol duration) spaced training sequence based time domain equalization. A 97-tap finite impulse response (FIR) filter is extracted from the training sequence with the taps updated by the recursive least square (RLS) algorithm. Then a Ts spaced decision-direct recursive least square (DD-RLS) filter is used to improve the signal quality. At last, a two-tap digital post filter [1, α] followed by MLSD is used to compensate for the system bandwidth limitation. Fig. 3(c) shows the frame structure of the transmitted Nyquist PAM-4 signal. The preamble includes two 64-symbol synchronization sequences and four 128-symbol training sequences. Following the preamble, 25600 data symbols are transmitted.

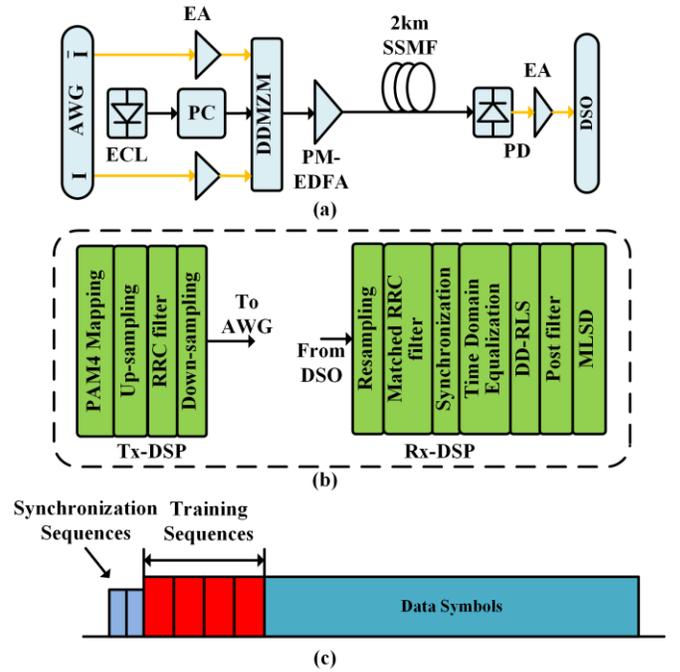

Fig. 3. (a) Experimental setup of PAM-4 transmission. (b) Tx-DSP and Rx-DSP algorithms. (c) Frame structure of transmitted signal.

### B. Result

Fig. 4(a) shows the measured BERs versus the received optical power at different scenarios. Without post filter and MLSD, neither the 2 km SSMF transmission nor back-to-back (BTB) scenario is possible to get a BER below the 7% HD-FEC threshold of $3.8 \times 10^{-3}$. With the help of post filter and MLSD, however, the BERs can be reduced by one order of magnitude, which are $5.55 \times 10^{-4}$ and $4.46 \times 10^{-4}$ for 2 km SSMF transmission and BTB scenarios, respectively. The net data rate is up to ~109 Gb/s with consideration of frame redundancy and 7% HD-FEC overhead. Fig. 4(b) shows the probability density profiles (PDFs) of signals in different transmission scenarios, which indicate that the received signal has just worsened a little after 2 km SSMF transmission. Our work is comparable to the PAM-4 transmission records based on silicon intensity MZMs that have been reported in [11, 12].

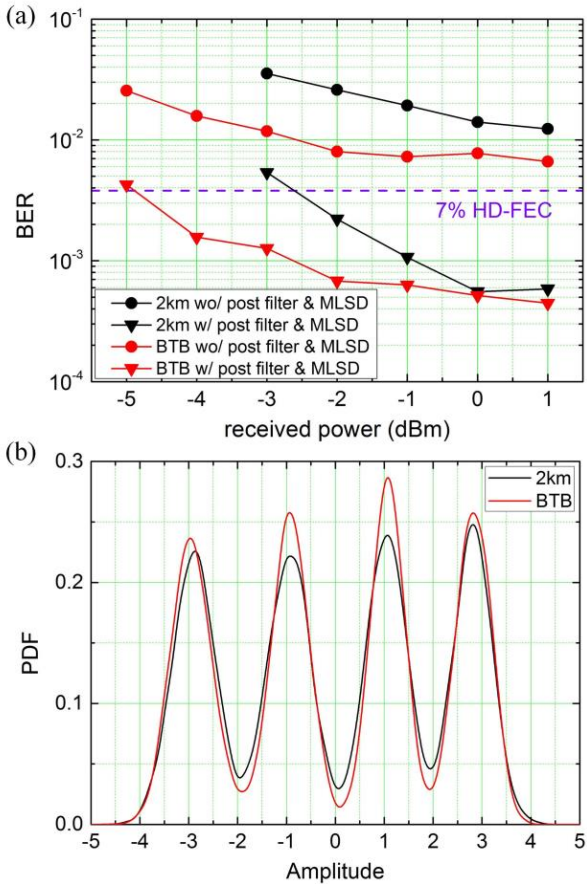

Fig. 4. (a) Measured BERs versus received power at different transmission scenarios with and without post filter and MLSD. (b) Probability density profiles (PDFs) in 2km transmission and BTB scenarios before post filter, respectively.

We also test the BERs at lower bitrates as shown in Fig. 5. For the signals with different baud rates, we optimize the tap coefficient of post filter to approach the signal shape before MLSD. Higher baud rate signal requires larger coefficient to suppress equalization-enhance noise in the high frequency region. For PAM-4 signals with bitrates of 110 Gb/s and 100 Gb/s, the BERs of $4.0 \times 10^{-5}$ and $2.0 \times 10^{-5}$ are achieved after 2 km SSMF, respectively, both of which are below the KP4 FEC threshold of $2.2 \times 10^{-4}$ [18].

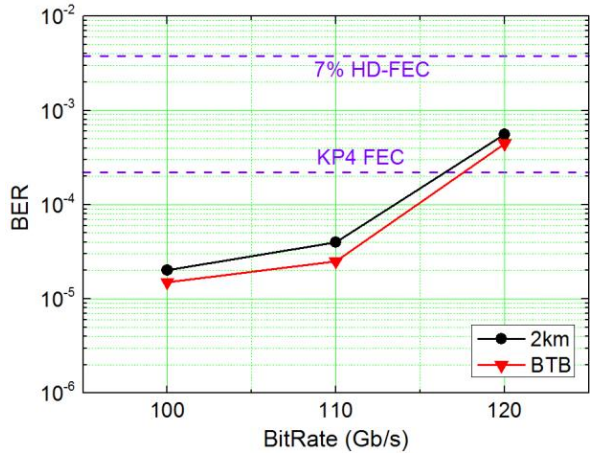

Fig. 5. Measured BERs as a function of bitrate in 2km transmission and BTB scenarios, respectively.

## IV. SSB PAM-4 Transmission

### A. Setup and DSP

Conventional PAM-4 modulation is limited in transmission distance because of the CD induced power fading in direct detection systems. We adopt SSB PAM-4 modulation to overcome this problem and meet the demand for metro-scale inter-DCIs. The same silicon DD-MZM is used as Section III.

The experimental setup for SSB PAM-4 signal is shown in Fig. 6(a), which is similar to Fig. 3(a) except that at the receiver side an EDFA is firstly used to compensate for the fiber loss and then an optical band-pass filter (OBPF) is added to suppress the out-of-band noise. It is worth noting that to generate SSB signal with a DD-MZM, working at the quadrature point is necessary to ensure a π/2 phase shift between the two arms of the MZM. Therefore, we set the ECL operating at 1552.9 nm, which is the same wavelength as PAM-4 experiment in Section III.

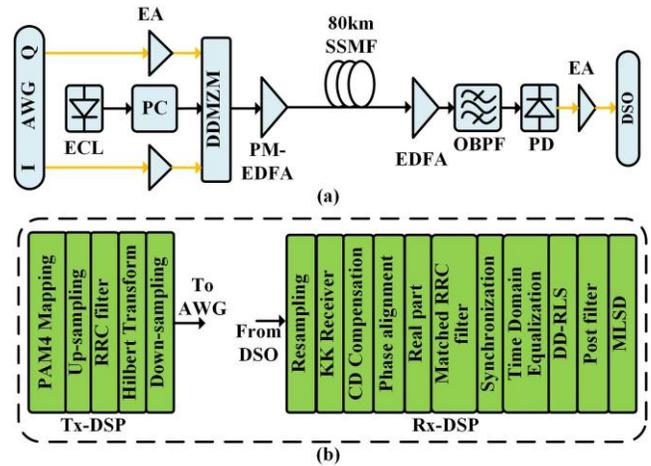

Fig. 6. (a) Experimental setup of SSB PAM-4 transmission. (b) Tx-DSP and Rx-DSP algorithms for SSB PAM-4 transmission.

The DSP diagram for SSB PAM-4 is shown in Fig. 6(b). At the transmitter, the bit stream is mapped to PAM-4 first. After 8 times up-sampling, the signal is digitally shaped using RRC filter with a roll-off factor of 0.01. Next, a digital Hilbert filter is applied to remove one of the two conjugated sidebands. Then the signal is 7 times down-sampled. Hence the baud rate is 56 Gbaud when the AWG is operated at 64 GSa/s.

For the receiver DSP, as the SSB signal would be converted to DSB signals after photodiode detection, 4 times resampling is performed to avoid aliasing due to the logarithm operation in KK detection. After the KK receiver and CD compensation, the PAM-4 signal is rotated with a constant phase shifter and then taken the real part to recover the original double-sideband signal [17]. The following DSP steps are similar to what have been mentioned in Fig. 3(c). The frame structure of the transmitted Nyquist SSB PAM-4 signal is the same as what has been shown in Fig. 3(c).

A theoretical description of SSB signal generation using DD-MZM has already been presented in [6]. The transfer function of a DD-MZM can be described as Eq. (1) using small signal approximation.

$$E_{out} \approx \frac{1}{2} E_{in} [\frac{\pi}{V_\pi}(V_I + jV_Q) + 1 - j]$$
$$= S + C\exp(-j\frac{\pi}{4}) \quad (1)$$

Here $V_I$ and $V_Q$ are independent RF signals driving the two arms of the DD-MZM, respectively, and the phase difference between the two arms is set to π/2 by using wavelength tuning. For SSB modulation, the driving signals applied to I and Q branches should be Hilbert transform pairs. From Eq. (1), it is also observed that the output optical field can be divided into two parts: the SSB signal and an extra carrier term with a phase rotation of -π/4.

Fig. 7 shows the signal spectra at different DSP stages at the receiver side. The constant phase rotation will be passed to the signal after PD detection as shown by the second term in Eq. (2), which cannot be removed by KK detection. Therefore, the phase alignment is required to shift the phase difference of the conjugated DSB signal from π/2 to 0.

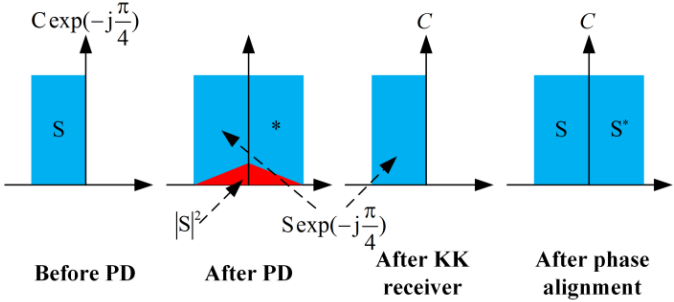

Fig. 7. Signal spectra at different DSP stages.

$$\left|C\exp(-j\frac{\pi}{4}) + S\right|^2 = |C|^2 + 2\operatorname{Re}\{C^* \cdot S\exp(j\frac{\pi}{4})\} + |S|^2 \quad (2)$$

*B. Result*

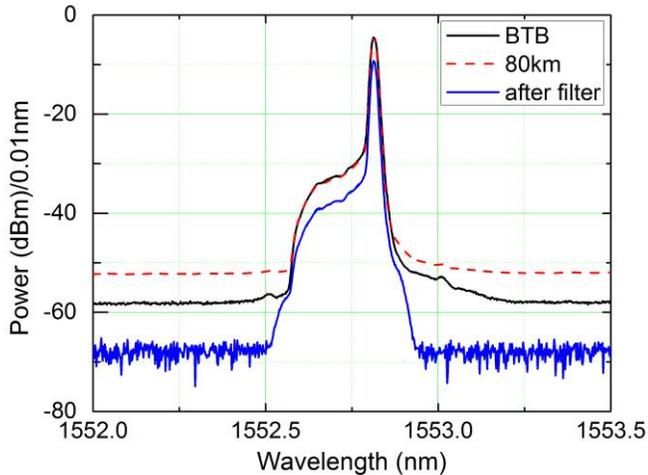

Fig. 8. Received optical spectra in different transmission scenarios with 0.01 nm resolution for SSB PAM-4.

Since SSB signals with larger optical sideband suppression ratio (OSSR) perform better in square-law detection, the time skew between AWG's two output channels should be first finely adjusted to maximize the OSSR. After OSSR maximization, we optimize the carrier-to-signal power ratio (CSPR) of the SSB signal. Once the bias voltages are fixed, the CSPR can only be adjusted by changing the RF driven voltage. Therefore, we drive the modulator with an appropriate RF driven voltage in order to generate SSB signal with an enough high signal to noise ratio (SNR) and meanwhile ensure a CSPR satisfying the minimum phase condition for KK detection [9]. Here the CSPR of the SSB PAM-4 signal is set to 16.6 dB. From the measured optical spectrum in BTB scenario shown in Fig. 8, we can see that the OSSR of the generated signal is nearly 20 dB, which indicates good signal quality. The filtered curve is also measured, which demonstrates that most of the out-of-band noise has been removed by the OBPF. The lower amplitude of filtered spectra is due to the insertion loss of the OBPF.

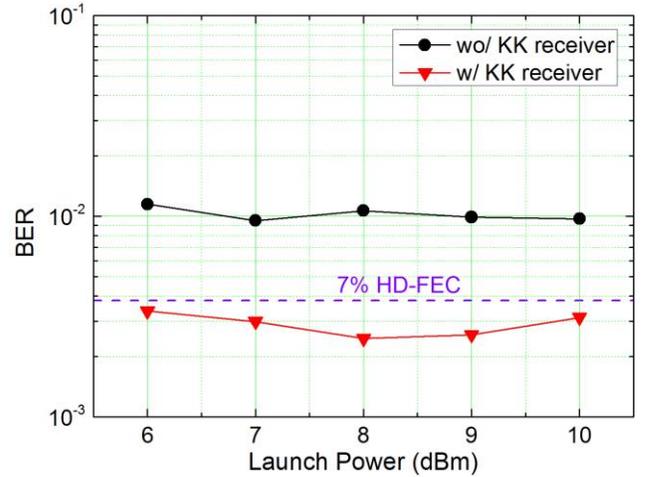

Fig. 9. Measured BERs as a function of launch power over 80 km SSMF transmission with/without KK receiver.

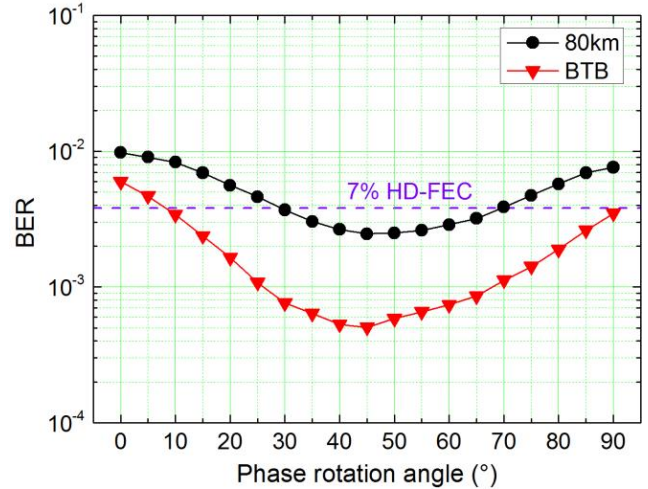

Fig. 10. BER versus phase rotation angle in the phase alignment operation for SSB PAM-4 signal in 80 km transmission and BTB scenarios, respectively.

Fig. 9 shows the BER as a function of total launch power for 80 km SSMF transmission. By using the KK receiver, the BER of SSB PAM-4 signal can be reduced below 7% HD-FEC threshold of $3.8 \times 10^{-3}$. With the KK receiver, the optimal total launch power is 8.0 dBm, and the optimal BER is $2.46 \times 10^{-3}$. The corresponding OSNR value is measured as 41.1 dB. We also optimize the phase rotation angle in the phase alignment operation. As shown in Fig. 10, the optimal phase rotation angle is 45° for both BTB and 80 km transmission scenarios, which coincides with the theoretical analysis above mentioned.

Fig. 11(a) shows the BER performance as a function of the optical signal-to-noise ratio (OSNR) for SSB PAM-4 signal at different transmission scenarios. The corresponding PDFs before the post filter are also shown in Fig. 11(b). The net bitrate of the signal is about 102 Gb/s with consideration of frame redundancy and 7% HD-FEC overhead. It can be observed that the OSNR penalty after 80 km transmission at the BER of $3.8 \times 10^{-3}$ is about 3.2 dB in comparison with the BTB result. The PDFs also illustrate that the SSB signal has not worsened significantly after 80 km SSMF transmission, which demonstrates that our SSB PAM-4 system is capable for metro-scale transmission.

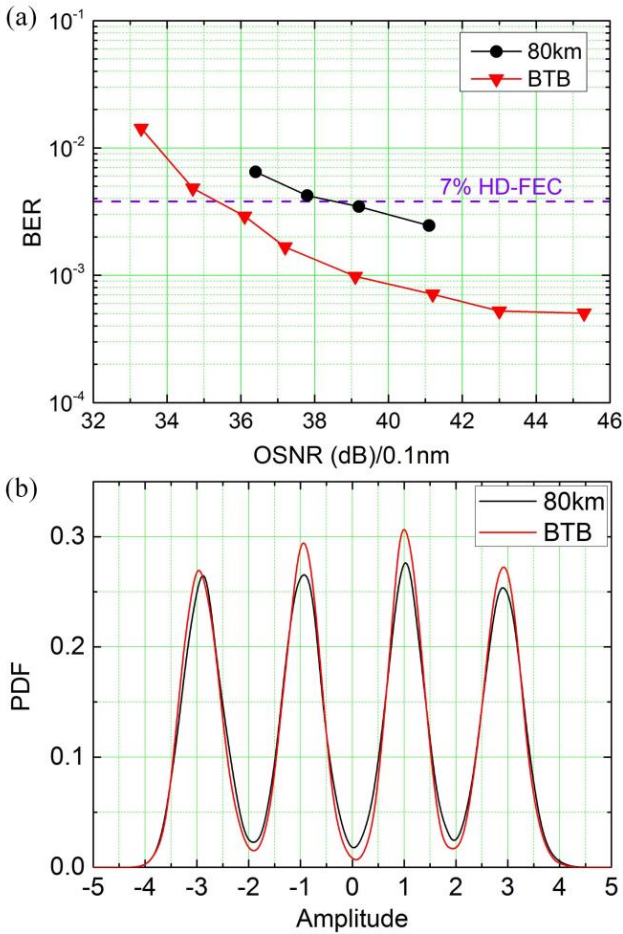

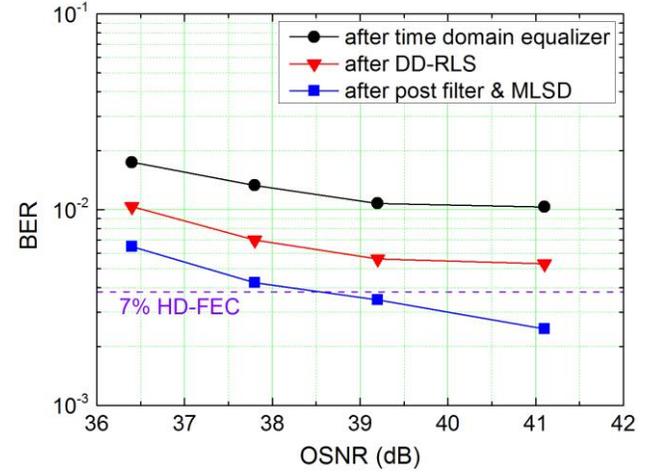

Fig. 12. Measured BERs as a function of OSNR for 80km transmission after different DSP steps.

We also compare the BER performance after different DSP steps in 80 km transmission scenario as shown in Fig. 12. It can be observed that after the first step of time domain equalization, the received BER is far above the 7% HD-FEC threshold. By using the DD-RLS filter, the signal quality is improved significantly and the BER is reduced. Finally, with the help of a two-tap digital post filter along with MLSD to mitigate the ISI, the signal quality is further promoted and the BER is below the 7% HD-FEC threshold. It can be concluded that the receiver side algorithms such as the DD-RLS filter and the post filter along with MLSD algorithm have played indispensable roles in bandwidth-limited high-speed SSB PAM-4 transmission.

Fig. 11. (a) Measured BERs as a function of OSNR for SSB PAM-4 signal in different transmission scenarios. (b) Probability density profiles (PDFs) of SSB PAM-4 signal before post filter in 80km transmission and BTB scenarios, respectively.

We have also tested BERs at a series of bitrates as shown in Fig. 13. For the signals with bitrates less than 112 Gb/s, we do not use post filter and MLSD anymore since they do not improve the BER performance for low baud rate. For SSB PAM-4 signals operating at 100 Gb/s, 88 Gb/s, 76 Gb/s, and 64 Gb/s, the BERs achieve $1.37 \times 10^{-3}$, $3.95 \times 10^{-4}$, $3.09 \times 10^{-4}$ and $1.07 \times 10^{-4}$, respectively, after 80 km transmission. It is worth noting that when the bitrate is reduced to 64 Gb/s, the BER of SSB PAM-4 signal after 80 km transmission is even below the KP4 FEC threshold of $2.2 \times 10^{-4}$. In our experiments, for the BER above $1 \times 10^{-4}$, $\sim 1 \times 10^{6}$ bits are used in error counting while for the BER lower than $1 \times 10^{-4}$, $\sim 1 \times 10^{7}$ bits are counted.

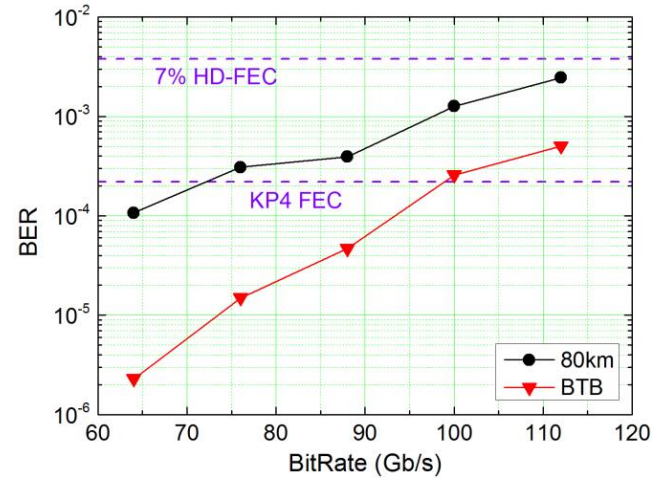

Fig. 13. Measured BERs as a function of bitrate for SSB PAM-4 signal in 80km transmission and BTB scenarios, respectively.

## V. CONCLUSION

In this paper, we fully explore the capability of a C-band silicon DD-MZM for data center interconnection applications with direct detection scheme. For short-reach intra-data center interconnections, we achieve 120 Gb/s PAM-4 transmission over 2 km SSMF with a BER only $5.55 \times 10^{-4}$, which is comparable to the best PAM-4 transmission records based on silicon depletion-type Mach-Zehnder modulator. DD-RLS filter is used to improve the signal quality. A two-tap digital post filter and MLSD are also used to compensate for the system bandwidth limitation. The net data rate is up to ~109 Gb/s with consideration of frame redundancy and the 7% HD-FEC overhead.

For metro-scale inter-data center interconnections, we successfully generate 56 Gbaud Nyquist SSB PAM-4 signal and demonstrate an experiment of 80 km SSMF C-band transmission. With the help of KK receiver, the DD-RLS filter and the post filter along with MLSD, the BER after transmission can be reduced to $2.46 \times 10^{-3}$, which is below the 7% HD-FEC threshold of $3.8 \times 10^{-3}$. The net bit rate of the system is about 102 Gb/s, to the best of our knowledge, which is also the highest single-lane bitrate for 80 km SSB signal transmission based on a silicon DD-MZM.

Our experiments show that due to its high performance, low cost and compact footprint, silicon DD-MZM is a competitive solution for high-speed optical transmission of both intra and inter data center interconnections.